# Flawing CERN antihydrogen-experiments with the available H-spectrum[1]

G. Van Hooydonk, Ghent University, Faculty of Sciences, Krijgslaan 281 S30, B-9000 Ghent (Belgium)

**Abstract**. Solving the H̲-problem could well be of historical interest but a solution must be unambiguous. We use already available and accurate spectral evidence to contradict and even to flaw the current CERN H̲-experiments, set up to unravel this H̲-mystery. Making H̲ with a *long-range interaction* between $e^+$ and $p^-$ is impossible, since this mass-asymmetrical pair of charge-conjugated antiparticles is confined to a bound state at *close-range*. This resembles the short- and long-range quark behavior in QCD. A real solution for H̲ will therefore require a different approach. Pacs: 34.10.+x, 34.90.+q, 36.10.-k

**Introduction**
Left-right asymmetry in naturally observed stable systems has a long history. With the seminal works of Pasteur in the 19th century, chiral behavior became of scientific interest. But only a century after chemists, physicists realized the importance of chirality for simple neutral particle systems. Today, there is a growing interest in the properties of H̲ for a variety of reasons [1]. Imaging an internal particle-structure for atom hydrogen seems simple. With charge conjugation symmetry **C**, H (particles $e^-,p^+$), *allowed in nature* is transformed in its mirrored version H̲ (antiparticles $e^+,p^-$), *forbidden in nature*. Consequently, one expects that neutral antimatter H̲ annihilates when it meets neutral matter but claims exist that mass-producing H̲ *seems* possible [2]. At CERN, one attempts to produce *artificial antihydrogen* H̲* [1,2] with combination reaction

$e^+ + p^- \quad \rightarrow \quad$ H̲*  (1)

which, at first sight, seems perfectly logical and valid beyond any doubt.
Known problems with interacting antiparticles are (a) the possibility that they show a different short- and long-range behavior, like quarks which are unstable at long but stable at short range, and (b) the effect of system symmetries. Perfectly symmetrical pairs of charge-conjugated particles, with the center of charges $c_C$ coinciding with the center of mass $c_M$ like in *positronium* or *antiprotonium*, are known to annihilate. But if $c_C$ does not coincide with $c_M$ like in (1), quasi-bound systems can be formed, where annihilation is, at least, delayed as demonstrated for instance with antiprotonic helium [3].
The CERN-based H̲-approaches [1,2] rely on the conventional belief that combination reaction (1) must lead to antihydrogen H̲ [1,2]. Reaction (1) *seems* justified since it is an exact mirror copy of the corresponding formation reaction for hydrogen H

$e^- + p^+ \quad \leftrightarrow \quad$ H  (2)

which is a natural, reversible and valid process. If (1) were really a valid consequence of (2) due to **C** and/or **P**, there is no need at all to question the work by ATHENA and ATRAP [1,2]. The trust of the physics community in the reliability of approach (1) is so great [1] that it is difficult to imagine different or alternative solution(s).
*But there is a case for which it can be proved with absolute certainty that reaction (1) is impossible: this is when natural antihydrogen H̲-states exist* [4-8]. This theoretical possibility explains why *artificial* antiH in (1) is denoted as H̲*, whereas natural hydrogen-states are referred to with symbol H̲. As we will show below, the crucial natural case H̲ can only be understood with a quark-like model.
Unlike for *artificial* H̲*, for which direct signatures have not yet been provided [1,2], the many concrete signatures for *natural* H̲-states are overwhelming: (a) evidence in the band spectrum of over 300 diatomic bonds, including molecular hydrogen in particular, which proves that intra-atomic charge-inversion occurs in nature [4,5], is indeed confirmed by (b) supporting *complementary* evidence in the line spectrum (Lyman-series) of natural atomic hydrogen [6-8]. Both *complementary signatures* provide with generic and system independent criteria for chiral behavior of

---

natural species hydrogen. In concreto, it is easy to derive a *Mexican hat curve* from the known spectrum of natural H to prove the chiral behavior of this simple neutral species [7,8]. An *inverse Mexican hat* curve is equivalent with a classical Van der waals-Maxwell-curve, probing and detailing the reversible natural H-H̲ phase transition in the so-called simple Coulomb electron-proton bond [8].

Whereas these latter signatures provide *direct* evidence for natural H̲-states [4-8], most of the evidence claimed for *artificially produced H̲\* (1) is indirect and necessarily based upon complicated annihilation patterns and their interpretation* [1,2].

Not even the slightest direct signature for H̲* has yet been presented by the ATHENA or ATRAP collaborations [1,2]. To identify positively a species in an experiment, only its spectral characteristics can provide conclusive evidence. This logic is strictly obeyed in standard analytical chemistry (see example [9]) and should apply equally well for CERN-H̲*-experiments. But the H̲*-spectrum is not yet known, since measuring this spectrum is exactly the main goal of [2], which could even make all claims on H̲*-production thus far reported [1,2] premature. In addition, Driscoll recently showed [10] that some of the preliminary results reported thus far perfectly fit in available theories for antiparticle plasma's.

Because of this *indirect* not-conclusive evidence for H̲* [2] and the *direct* conclusive evidence for H̲ [4-8], the possibility even remains *that H̲\* was not produced at all* [9], *mainly due to annihilation on the spot*. In essence, all claims [2] are *based upon the general conventional belief that (1) is plausible when compared with (2)* but a rigorous proof for the absolute validity of (1) was never given.

**Natural or artificial antihydrogen: H̲ or H̲***

With direct evidence [4-8], neutral species hydrogen can exist in *2 mutually exclusive quantum states*, a natural atomic hydrogen or H- and an equally natural but *charge-inverted* anti-atomic antihydrogen H̲-state. A reversible phase transition between these 2 left- and right-handed states can be described schematically as

$H_L$ (H̲) ↔ [$H_{crit}$] ↔ $H_R$ (H)                                    (3)

*an internal phase transition in a bound state* between enantiomers $H_L$ (H̲) and $H_R$ (H). This left-right distinction or the *chiral behavior of natural species hydrogen is due to charge-inversion* **C** [4,5], which is also the result of applying **P** to the atomic dipole. Reversible transition (3) can only be triggered by a field-effect (heat, radiation…), i.e. by a perturbation of *achiral* intermediary and critical state $H_{crit}$ in between $H_L$ and $H_R$.

Chirality related process (3) also has a long history as a similar transition applies for chiral behavior of real stable structures (enantiomers), discovered by Pasteur in the 19$^{th}$ century. A classical stable chemical left handed structure $ABCD_L$ can go over, in a continuous manner, into a right handed structure $ABCD_R$, through a process described accurately with CCMs, *continous chirality measures* [10]. In between the two enantiomers L and R, a critical, non-chiral or achiral state $ABCD_{crit}$ appears, exactly as indicated in (3) [11].

With generic scheme (3) based upon [4-8], flawing (1) is possible, pending the order of states H̲ and H with respect to $H_{crit}$. In fact, $H_R$ or atomic H is known to dissociate in an electron and a proton, as argued with classical and valid equation (2). But with (3), allowing electron and proton to get closer and closer, they will have to adapt to form *critical achiral configuration* $H_{crit}$, the result of a perturbation of the 2 unit charge Bohr H-structure. Further compression beyond achiral $H_{crit}$ can provoke *a permutation of the 2 unit charges* to give a different handedness for *a very closely bound and stable positron-antiproton system*, denoted by H̲ (or $H_L$) in (3).

If the classification (order) of mutually exclusive natural states H and H̲ in (3) is proved to be in line with observed reaction (2), (1) must be invalidated immediately. With mutually exclusive states ordered exactly as in (3), $H_L$ (or *natural* H̲) *can never dissociate directly in* $e^+ + p^-$, which makes reverse combination reaction (1) impossible indeed for any natural and stable H̲-state. Before dissociating, $H_L$ (or *natural* H̲) must first adapt according to *critical achiral* $H_{crit}$, which means its charges must invert as in normal Bohr-like H. If so, (1) used in [2] can never lead to *natural* H̲ directly: $H_{crit}$ is the barrier, which prevents reaction (1) to be possible[2] and simply contradicts any claim based upon [2].

It appears that, with natural H-states, approaches (1) and (3) are mutually exclusive too, since, if one of the two is valid, the other must be excluded: only one anti-atomic species $\underline{H}(e^+,p^-)$ can exist with intra-atomic charge inversion as it is confined to short-range only. As a result, if (1) leads to something like $\underline{H}^*$ as claimed by [2], this artificial structure can never be identical with natural $\underline{H}$.

An important difficulty with (1) is that it is just the result of a general convention, rather than of scientific reasoning. As a matter of fact, mainly *by convention, only the artificial CERN-based approach (1) is allowed in nature, while more natural (3) is excluded, despite the abundant spectroscopic evidence in its favor* [4-8].

Strictly spoken, it is impossible, without explicit proof, to forbid one of the two schemes (1) and (3) a priori. But a simple and straightforward phenomenological analysis of the H-line spectrum (the Lyman series) fixes critical *achiral* state $H_{crit}$ at $n_{crit}=\pi$ [6-8]. The other *chiral* states are classified easily: H-states are confined to domain $n>\pi$, in line with (2), whereas $\underline{H}$-states must be to the left of $H_{crit}$ as in (3), where $n<\pi$ [6-8]. Critical n-values related to number $\pi$ are consistent with de Broglie's standing wave equation [6]. Model (3) for the reversible $\underline{H}$-H transition is confirmed experimentally by the available line spectrum of natural hydrogen [6-8]. The intra-atomic charge inversion itself is proved with the band spectrum of molecular hydrogen [5]. As a result, (1) is impossible, because the existence of a critical intermediate achiral complex $H_{crit}$ is not accounted for in the CERN-approach [2].

Only a reversal of the order of $\underline{H}$- and H-states in (3) could come to the rescue for (1), since this would make reaction (1) allowed. But reversing the order of mutually exclusive states leads to meaningless results, rebutted by the logic behind scientific metrology and observation and by pure common sense, as is easily proved with two examples. The oldest one is even pre-quantal, since it is the classical macroscopic phase transition between two ordered or classified different states of aggregation of the same species [8].

**Reversing the order of two mutually exclusive states is always forbidden**
Let us discuss some consequences of *ordering mutually exclusive states or phases*. For 2 mutually exclusive states or phases to exist, they must be distinguishable in a field, say on a 1D semi-axis, for if not, they are degenerate. A distinction of states by a field implies an order, a classification.
(a) A Schrödinger cat has 2 mutually exclusive states, *dead* and *alive*, since, by definition, *alive = not dead*. This excludes the possibility that a cat is partially dead and partially alive or, which is equivalent, that a cat is in both states at the same time. *Only a transition from completely alive to completely dead is an allowed phase transition for a cat*, seemingly similar to (3). On the field or time semi-axis, the transition from alive to dead is allowed with increasing time *but only in this order* (irreversible transition). The inverse transition (dead to alive) for a single cat on the same time semi-axis is forbidden. On semi-axis +t, a permutation (inversion) of marks *alive* and *dead* for a single cat is not allowed, as it does not make sense.

Inverting the order of the marks alive and dead on the +t semi-axis would lead to stupid results: *a living man, looking back in time, would see he was already dead.*
(b) In a *reversible* phase transition, the order of states is as important as in (a). Let us replace the semi-axis +t by a volume axis +V (the field effect) and look at a macroscopic gas-liquid transition with the Van der waals-equation of state (EOS) [8]. Gas and liquid states for a single species are as mutually exclusive as dead and alive for a cat or as two discrete quantum states. Even in a mixture, each individual molecule must belong either to the liquid or to the gaseous state, as it cannot be in two different states at the same time. During a liquid-gas phase transition, the main properties of the unit species, say molecule H₂O, are not affected. While transforming from gas to liquid with the two phases temporarily in coexistence, the characteristic temperature (100 °C) or the pressure in the P,V-diagram cannot change, according to the Gibbs phase rule. It is trivial however that *at very large V, the system is normally in the gaseous state. Only when V decreases by compression*, the gas-liquid transition can set in and eventually lead to the liquid phase for the complete system, *whereby V for the liquid is always much smaller than for the gas. Also here, the domains for gas and liquid in the P, V-diagram can never be inverted. The borderline is determined by the intermediary maximum pressure $P_{crit}$ in the P, V-diagram* [8]. *This critical maximum distinguishes between a domain where the pressure increases (+) and one where the pressure decreases (-), eventually to end in a domain of negative pressures, difficult to understand with classical physics.*

---
[2] Except for tunneling effects, not to be discussed at this instance

Exactly as in (a), inverting marks for gas and liquid on the +V-axis leads to the same stupid results: *with an inversion of states of aggregation, one would obtain a gas by compressing (or cooling) a liquid. With these trivial examples, it is obvious that, marks or domains attributed to 2 mutually exclusive states or phases for the same species can never be inverted on the same field axis. Once the properties of its states or phases on the field axis have been defined by measurement or observation, their permutation is not allowed.*

This explains why the 2 mutually exclusive states for species hydrogen in (3) cannot be inverted either. As a consequence of the same logic, (1) would be impossible indeed.

**Mutually exclusive hydrogen- and antihydrogen-states on the field axis**

These trivial but stringent examples are further quantified with a typical Van der waals-curve focusing on the phase-transition itself, as shown in Fig. 1. The P, V-curve has an intermediary critical maximum $P_{crit}$, corresponding with a critical volume $V_{crit}$, important for the gas-liquid transition. The borderline between gas and liquid domains contains point $+P_{crit}$, crosses the +V-axis at $+V_{crit}$ and runs parallel with the P-axis as indicated in Fig. 1. It is evident that, once the +V semi-axis is fixed, the permutation of gas ($V_R > V_{crit}$) and liquid ($V_L < V_{crit}$) domains is forbidden. With model (3), the liquid-gas phase transition on the +V semi axis obeys the order

$$V_L \text{ (liquid)} \quad \leftrightarrow \quad [V_{crit}] \quad \leftrightarrow \quad V_R \text{ (gas)} \tag{4}$$

In Fig. 2, we show the generic schemes (without the quantitative fine structure of Fig. 1) for the +t semi-axis and the mutually exclusive cat states alive and dead. A similar scheme applies for model (3) with the +n semi axis to describe the states available for atomic species hydrogen. If n is Bohr's principal quantum number, *large n or $n > n_{crit}$ means a large separation for electron and proton* (pseudo-gaseous state, *low particle density*). When n decreases, electron and proton will form the conventional Bohr H-atom (*higher particle density*), according to (2) and in agreement with observation. In terms of liquid and gas, the Bohr-atom must be placed at the right-hand side of $H_{crit}$ with $n_{crit} = \pi$ (playing the role of $P_{crit}$ and $V_{crit}$ for liquid and gas in (4) and in Fig. 1). *Expanding the Bohr H-atom (increasing n further) cannot but lead to $e^- + p^+$*, in agreement with standard reaction (2). As H̲ is confined to $n < n_{crit}$ (see above), (1) is flawed as suggested in the preceding section, unless

(i) H̲ is indeed different from H̲* (see above) or *unless*
(ii) *Fig. 1 does not apply to chiral model (3) for natural hydrogen.*

**Why CERN H̲*-experiments must be flawed immediately**

It is obvious that something very elementary seems wrong with the finer quantitative details of the Van der waals-curve as depicted in Fig. 1. In fact, *it is impossible to measure negative pressures.* According to Maxwell, the surface described by the positive pressure domain, above the condensation line, must match exactly that by the negative pressure domain below it. *Classically, the data points drawn in Fig. 1 are not assessable by experiment* but, as shown elsewhere [8], the data points given explicitly in Fig. 1 are actually measured with relatively great spectroscopic accuracy. The underlying reason is that macroscopic energies PV correspond with energies of type $1/r$ if the force is of type $1/r^2$ [8]. Replacing the volume V by an inter-particle separation r for a Coulomb attraction $-e^2/r$ between 2 charge-conjugated particles like electron and proton and the perturbing field kT by hν=hc/λ, leads to the same logic for a lepton-baryon system disturbed by radiation as that for a macroscopic liquid-gas system, provided there is a critical separation at $n_{crit}$, corresponding with $H_{crit}$ in (3). So the quantitative picture generated by the Van der waals-equation for a macroscopic phase transition (4) can equally well apply for the microscopic phase transition in a neutral quantum system (3) [8], as remarked in the Introduction.

*But the most remarkable thing about the fine structure given in Fig. 1 is that all the data points on the curve therein are simply extracted from the Lyman series of natural atom H, available for many a decade* [7,8]. *In fact, the curve is the inverted Mexican hat potential, detected previously [7,8]. The subtle difference between the two is a representation of the data, either plotted versus 1/n (giving a Mexican hat curve) or versus n (giving a Van der waals-curve) and a very small asymptote shift* [8].

Analytically, this difference transforms $n_{crit}=\pi$ in the Mexican hat potential in a critical state for n between 5 and 6, as indicated in Fig. 1, the further details of which are given in [8].

The immediate consequence is that the natural H-domain is now proved to be confined to $n<n_{crit}$ and to the left of $H_{crit}$, as indicated in (3). With the same logic of the examples (a) and (b), inverting the marks for H and H on the +n-axis is not allowed as it would lead to equally stupid results. The domain for natural H is confined $n<n_{crit}$, with $n_{crit}=\pi$ for achiral state $H_{crit}$ in (3) in the Mexican hat representation of [7].

If we were to *expand* natural H from $n<n_{crit}$ by increasing n, the system reorganizes until it first reaches $n_{crit}$ or critical achiral state $H_{crit}$. After passing that critical configuration, it has transformed (*by charge-inversion* [5]) in natural Bohr-like H, which after further expansion, produces conventional result electron $e^-$ and proton $p^+$, exactly as described by natural process (2).

With (3) and the Mexican hat potential [7], it is even a simple matter to determine the energy of the critical state $H_{crit}$. If the harmonic Rydberg [6,7] is used as asymptote, the value for the maximum at $n_{crit}=\pi$ separating the two wells is

$$\Delta E(H_{crit}) = 0{,}044667 \text{ cm}^{-1} \qquad (5)$$

(error of order $10^{-5}$). Detailed information of this type for H, not yet available and even unthinkable with [2], proves why our results indeed *run ahead of the CERN experiments*, as argued before [6,7]. When the Bohr ground state is used as the asymptote, the perfect Mexican hat curve is distorted, producing an extreme for n between 5 and 6 where the depth of the well for H-states is about 0,01 cm$^{-1}$ [8] as indicated in Fig. 1. These shifts in asymptotes (asymptotic freedom) and in the critical points are easily assessable since, with chiral symmetry breaking, the levels in the H-Lyman $E_n$ series are given by

$$-E_n = R_{harm}/n^2 - A(1 - \tfrac{1}{2}\pi/n)^2/n^2 \qquad \text{cm}^{-1} \qquad (6)$$

which also gives a *chiral* explanation for the observed Lamb shift [6]. With (6), the quantitative relation between critical n and an asymptote shift is obtained analytically [8].

The direct implication of the above for CERN-approach (1) is that the marks for mutually exclusive states H and H on the +n semi-axis have *unjustly* been reversed (a permutation), which is not allowed by virtue of the common sense examples above. CERN's artificial H* is a state not allowed in nature, as, unlike natural H, H* is not ordered on the field axis as it should. *In short, trying to work out reaction (1) as in the highly praised CERN experiments [2] is as impossible as trying to make a gas by condensing or compressing a liquid.* This is why, on the basis of the available line spectrum of natural species hydrogen, CERN-approach (1) [2] must be flawed immediately. With respect to spectral evidence and history, it appears that (1) is based upon an inadequate if not erratic interpretation of available spectra for both molecular and atomic hydrogen [4-8].

**QCD, quarks, antiparticles and antihydrogen. Nuclear stability**

Although some problems with H can be solved with the already available spectra [4-8] and without the complicated experiments [1,2], a new problem appears: the interpretation of natural intermediary bound and *achiral* state $H_{crit}$, i.e. barrier (5), which prevents reaction (1) to be possible. It is evident that, if H has charge distribution +1, H has charge-inverted distribution −1, while both are still Coulomb systems. But this also means that bound state $H_{crit}$ must somehow have *zero charge distribution* or that *lepton and baryon or anti-lepton and anti-baryon are bound at close range but no longer exclusively by the Coulomb law.* This moderate part of the stability (energy) of the complete system is exactly the part described by the errors of *achiral* Bohr theory [6,7]. This must not come as a surprise since, in all chiral structures observed in nature, the left-right difference does not have a drastic effect on its stability (energy). *Only the left-right morphology of chiral structures is affected, be it in a mutually exclusive way.*

To understand this very same small fraction of the total H-energy beyond Bohr H-theory, QED was developed. This is essentially Dirac theory in the Coulomb field but the connection with chiral behavior of H was never made [6].

In bound state $H_{crit}$, particle pair electron-proton as well as antiparticle pair positron-antiproton at short range do no longer interact exclusively by means Coulomb law, although they are known to interact strongly at long range, *which is typical for quark behavior*. A symmetric or achiral state in molecular hydrogen has the same effect on 4 lepton-baryon Coulomb interactions, which vanish from the scene exactly by virtue of a very specific structural symmetry [5]. An important byproduct of our analysis of atomic and of molecular hydrogen as well is *the important effect of asymptotic freedom exactly as it is for QCD* [5,8]. We conclude therefore that a quark model like QCD is needed to understand atomic and molecular hydrogen, *the 2 simplest bound elementary particle systems*, *but readily assessable by spectroscopic methods* [4,5].

A seemingly speculative but tempting solution for most of these problems with H is to explore the connection between quarks and antiparticles. Both are essential to understand bound states at short range but they cannot be isolated. At long-range, it proves impossible to detect or to isolate quarks or antiparticles as individual particles.

Reconciling the fractional charge of quarks with the unit charge of antiparticles is easily done, as explained in [5]. Quite surprisingly, the solution for this charge problem has its roots, exactly like the Van der waals-curve, in the 19[th] century [4-8].

An additional common sense argument in favor of natural intra-atomic charge-inversions is, obviously, the stability of (higher Z-) nuclei. Adhering to the same concept as above for Z=1 as in scheme (3), it can be expected that, at long range, the conventional description holds for +Z> +1, with its repulsive Coulomb character of order $+¼Z^2/r$. At short range, below some critical r, an antiparticle configuration will (gradually) appear. Then, in the intermediate *achiral* nuclear state corresponding with scheme (3) for H, reached when half of the protons turn into antiprotons, *Coulomb repulsion* transforms in *Coulomb attraction* of order $-¼Z^2/r$. This may remove the conventional Coulomb repulsion at the nuclear level [4], which led, amongst others, to the quark model above. Next, a simple explanation is given for the apparent difference between even-Z and odd-Z nuclei.

**Physical difference between artificial H\* and natural H**

We showed that the distinction between *natural* H and H-states may refer to reduced mass μ for system H, although absolute mass $m_H$ remains unchanged under the transition (3). So, natural mutually exclusive states H and H are separated by an internal algebra (parity) in reduced mass

$$\mu_\pm = m_e(1 \pm m_e/m_H) \tag{7}$$

as in [6,12] but invisible in bound state QED [13]. The minus solution of (7) gives the reduced mass in Bohr- and QED H-theories. Artificial CERN version H\* is connected with this same reduced mass by (1), but, in reality, inappropriate for antiH. This explains why H\* and (1) are not allowed in nature and why (1) is impossible. The reduced mass for natural H is connected with the positive sign in (7) used in Bohr theory [6]. With (7), the ratio of reduced masses for natural H and H is of the order 1,0011. This is close to the observed anomalous mass of the *free* electron, i.e. 1,001159, a strange coincidence indeed [6,12].

Details of a 4-fermion chiral complex for perturbed hydrogen by radiation, as in (3), are in [14]. This brings (3) more in line with classical 19[th] century chiral behavior, as it should [14].

**Conclusion**

A Van der waals-curve hidden in the line spectrum of atom hydrogen proves that scheme (3) is valid, that natural H-states exist [3-6] and that, with formation reaction (1) used by ATHENA and ATRAP collaborations [2], *it is impossible to manufacture natural H*. Understanding critical symmetrical or achiral bound state of atomic hydrogen, seemingly with zero charges and with relative energy (5), requires a QCD-like approach for 2 unit charge, neutral atomic hydrogen. In this context, the importance of an algebraic reduced mass (7) cannot be underestimated.

As far as history is concerned, poor communication between physicists and chemists on the interpretation of available spectral data for N-unit charge systems like atomic and molecular hydrogen (the prototype chemical bond) is at the roots of unjust working hypothesis (1), the basis of [2]. *Maybe, working out the connection between quarks and antiparticles, both needed to explain the chiral behavior of bound stable composite particles, can lead to a highly desirable common and more unifying solution.*

**Acknowledgments.** I am in debt to M. Tomaselli (GSI, Darmstadt) for discussions.

Fig. 1 Details of a macroscopic phase transition

    with the typical Van der waals P,V-curve and the critical point

    for pressure and volume

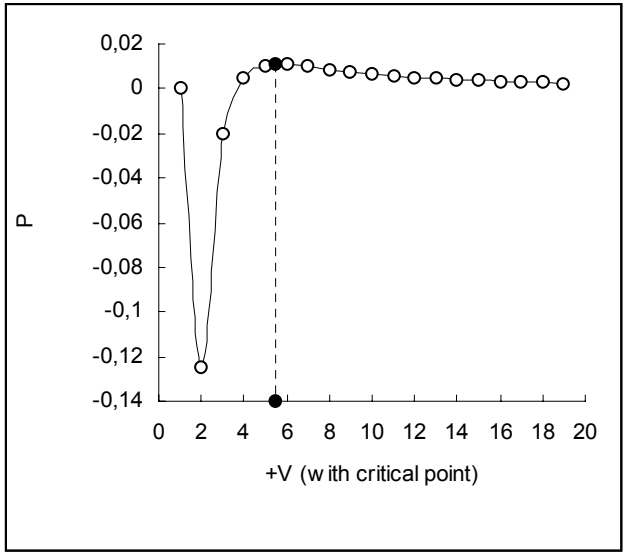

Fig. 2 Schema for parity-related mutually exclusive states +1 and –1:

    states dead and alive (+t-axis) and

    states antihydrogen and hydrogen (+n-axis)

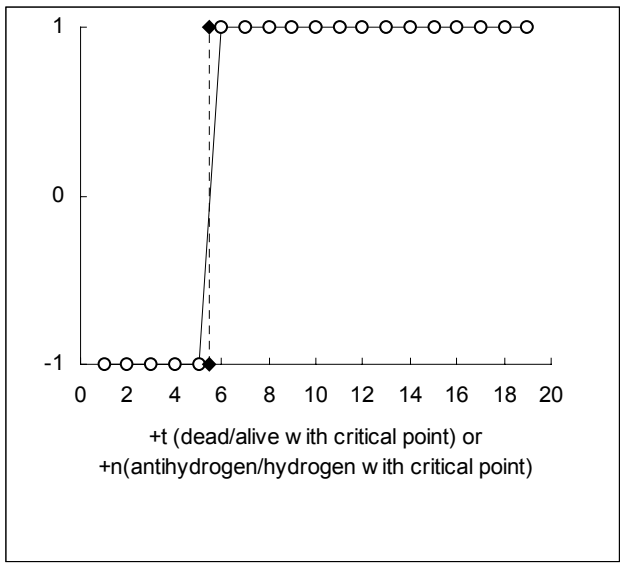